%Paper: hep-ph/9211319
%From: <BOS%UCLAHEP.BITNET@CORNELLC.cit.cornell.edu>
%Date: Wed, 25 Nov 92 18:17 PDT

% Explicit Calculation of the Renormalized Singlet Axial Anomaly,
% M. Bos, harvmac, 28 pages plus 4 tables (figures not included,
% available upon request), UCLA/92/TEP/41

\input harvmac

\Title{UCLA/92/TEP/41}{\vbox{\centerline{Explicit Calculation of the
Renormalized}
\vskip2pt\centerline{Singlet Axial Anomaly}}}

\centerline{Michiel Bos\footnote{$^*$}
{e-mail: bos@uclahep.jnet, or bos@uclahep.physics.ucla.edu.}}
\bigskip\centerline{Department of Physics}
\centerline{University of California}\centerline{Los Angeles, CA 90024}
\vskip .3in

A careful and complete discussion is given of the renormalization
of the singlet axial anomaly equation in a vector-like nonabelian
gauge theory such as QCD regularized by ordinary dimensional
regularization. Pseudotensorial structures are treated with
the 't Hooft-Veltman prescription. A general framework for calculations
is developed, and subsequently verified by explicit computations through two
loops. This is followed by a discussion of the matrix elements
obtained.

\Date{11/92}

\hoffset=.25truein
\voffset=.25truein
\hsize=6truein
\vsize=8.5truein
\baselineskip=16pt

\def\Dsl{\,\raise.15ex\hbox{/}\mkern-13.5mu D}
\def\dsl{\raise.15ex\hbox{/}\kern-.57em\partial}
\def\psl{{\raise.15ex\hbox{/}\kern-.57em {p}}}
\def\darr#1{\raise1.5ex\hbox{$\leftrightarrow$}\mkern-16.5mu #1}
\def\larr#1{\raise1.5ex\hbox{$\leftarrow$}\mkern-16.5mu #1}
\def\FFd{\epsilon^{\mu\nu\rho\sigma}F^a_{\mu\nu}F^a_{\rho\sigma}}
\def\ffd{{F\tilde F}}

\def\ffdc{{[\ffd]_C}}
\def\ffdms{{[F\tilde F]_{MS}}}
\def\ffdos{{[F\tilde F]}_{OS}}
\def\jms{{[j_5]_{MS}}}
\def\jos{{[j_5]_{OS}}}
\def\1L{\vert_{{one\>\> loop}}}
\def\2L{\vert_{{two\>\> loops}}}
\def\Gh{\hat\Gamma}

\def\mjms{{m[j_5]_{MS}}}
\def\mjos{{m[j_5]_{OS}}}
\def\dj{{\partial_\mu j^\mu_5}}
\def\djms{{\partial_\mu [j^\mu_5]_{MS}}}
\def\djos{{\partial_\mu {[j^\mu_5]}_{OS}}}
\def\tr{{\rm tr}}
\def\O{{\cal O}}
\def\Oan{\O_{an}}
\def\Oct{\O_{ct}}
\def\is{\! = \!}
\def\isnot{\!\not=\!}
\def\gap{\>\>\>\>\>}
\def\1L{\vert_{{one\>\> loop}}}
\def\2L{\vert_{{two\>\> loops}}}
\def\Gh{\hat\Gamma}
\def\to{\!\rightarrow\!}

\newsec{Introduction}

\nref\ABJ{J.~Bell and R.~Jackiw, Nuovo Cim. 60A, 47 (1969);
S.~Adler, Phys. Rev. 177, 2426.}
\nref\Oldan{H.~Fukuda and Y.~Miyamoto, Prog. Theor. Phys. 4, 347 (1949);
J.~Steinberger, Phys Rev. 76, 1180 (1949); J.~Schwinger, Phys. Rev.
82, 664 (1951); K.~Johnson, Phys. Lett. 5, 253 (1963).}
\nref\TJZW{S.~Treiman, R.~Jackiw, B.~Zumino, and E.~Witten, {\it Current
Algebra and Anomalies} (Princeton University Press, Princeton, 1985).}
The axial anomaly \refs{\ABJ,\Oldan,\TJZW}\ is by now a standard
feature of gauge field theory. In spite of this the study of its
renormalization has still not reached a state of stable equilibrium.
The literature on the question, which began more than twenty years ago
with the classical
work of Adler and Bardeen \ref\AB{S.~Adler and W.~Bardeen, Phys. Rev. 182,
1517 (1969); S.~Adler, in {\it Lectures on Elementary Particles and
Quantum Field Theory}, ed. S.~Deser, M.~Grisaru and
H.~Pendleton (MIT Press, Cambridge Mass. 1970).},
has grown both by the emergence of new issues and by the reconsideration of
old ones, a process that continues to the present day. Particularly notable
revivals of interest have been spurred by the paradox of the
currents in supersymmetric theories (see e.g. \ref\SUSY{M.~Grisaru,
in {\it Recent Developments in Gravitation}, ed. M.~Levy and S.~Deser
(Plenum Press, New York, 1979).}\ for an early formulation of the
problem)
and, more recently, by new developments concerning the spin of the
proton (reviewed e.g. in \ref\Alt{G.~Altarelli, in {\it The Challenging
Questions}, Proc. of the 1989 Erice Summer School, ed. A.~Zichichi
(Plenum, New York, 1990).}).

The present paper intends to contribute to the topic an explicit
perturbative computation at the two-loop level for a vector-like
nonabelian gauge theory such as QCD. More precisely, I will calculate the
two-gluon matrix elements of all the operators in the singlet anomaly
equation to second nontrivial order in dimensionally regularized
perturbation theory, and apply the results to an analysis of
the equation itself. The choice of ordinary dimensional regularization
(as against the dimensional reduction of much of the
supersymmetric literature, e.g. \ref\GMZ{M.~Grisaru, B.~Milewski and
D.~Zanon, Nucl. Phys. B266, 589 (1986).}) is based on its prominence in
perturbative studies of nonabelian gauge theories, as well as its
relatively obvious consistency. The prescription of 't Hooft and
Veltman \ref\tHV{G.~'t~Hooft and M.~Veltman, Nucl. Phys. B44, 189
(1972).} is adopted for the pseudotensorial objects
$\gamma_5$ and $\epsilon^{\mu\nu\rho\sigma}$. This type of approach is
of course not novel in itself; the present investigation
is similar in spirit to several earlier ones, notably
\ref\JL{D.R.T.~Jones and J.P.~Leveille, Nucl. Phys. B206, 473 (1982).}
and \ref\Akh{R.~Akhoury and S.~Titard, Michigan/Southhampton preprint
UM-TH-91-21, SHEP-91/92-2 (1991).}. I have gone beyond this earlier work
by performing calculations that are both more comprehensive and more complete;
I have also gone to greater lengths in connecting the actual computations
to their conceptual underpinnings.

The first two sections following this introduction are devoted
to a general derivation of the renormalized anomaly equation. First
I review the operators of interest, normalizing them with reference
to their scaling properties to the extent allowed by universality;
renormalization group invariance of a renormalized anomaly equation
then leads to strong a posteriori constraints on its
coefficients. This line of argument, which originated in early
work on the intertwining of the anomaly and Callan-Symanzik equations
\nref\Zee{A.~Zee, Phys. Rev. Lett. 29, 1198 (1972).}\nref\PSh{S.-Y.~Pi
and S.-S.~Shei, Phys. Rev. D11, 2946 (1975).}\refs{\Zee,\PSh},
has by now become rather commonplace in the literature,
where one finds it expounded with varying degrees of precision.
Next I adopt a constructive approach, regularizing the theory via
dimensional regularization, deriving the anomaly
equation and renormalizing it, and matching the result with the
scheme-independent template obtained before.

Section 4, which is the core of the paper, contains most of
the actual calculations performed. Apart from the very validity
of the renormalized anomaly equation they verify the
finite renormalizations required in the formalism of dimensional
regularization. In view of the role of renormalization group arguments in
organizing the problem I have employed the background field method,
within which the renormalization of the nonabelian theory is most
analogous to that of the well-understood abelian case.
I pay particular attention to the normalization
of the gluon tensor, which has been taken for granted in
previous treatments along similar lines.

The final section is devoted to comments on the matrix elements
obtained, mostly from points of view suggested by the abelian
Adler-Bardeen theorem \AB. The discussion eventually touches on an ongoing
discussion about a possible renormalization of the vacuum angle
in QCD \nref\SVI{M.A.~Shifman and A.I.~Vainshtein, Nucl. Phys. B277,
456 (1986).}\nref\SVII{M.A.~Shifman and A.I.~Vainshtein, Nucl. Phys. B365,
312 (1991).}\nref\Jog{A.~Jogansen, Sov. J. Nucl. Phys. 54, 349 (1991).}
\refs{\SVI,\SVII,\Jog}.

The literature on anomalies is so extensive as to preclude
exhaustive citation. My list of references is a selection of the
papers best known to me, and I apologize to authors
of omitted work.

The system considered consists of a generically nonabelian gauge field $A$
with coupling constant $g^2$, and fermions $\psi$ of mass $m$ forming
$n_R$ copies of an irreducible representation $R$ of the gauge group.
The group invariants $c_R$ and $T_R$ are defined in the usual manner,
$\tr\> T^aT^b=T_R\delta_{ab}$, $T^aT^a=c_R{\bf 1}$, where the $T^a$
are the group generators, $\bigl[T^a,T^b\bigr]=if_{abc}T^c$; in
particular, for the adjoint representation $c_A\delta_{ab}=
f_{acd}f_{bcd}$. I denote the dimensionality of the regulating
spacetime of dimensional regularization by $d$ and write $d=4-\epsilon$.

\newsec{Operator Normalization}

This section reviews the general structure of the
anomaly equation. The theory is
assumed to be regularized in some gauge-invariant way, but the details
of the regularization will be irrelevant. The point is to describe
a sensible normalization of relevant operators by
concentrating on scaling behavior and universality arguments.

The following auxiliary consideration will be required.
Suppose one is given a renormalized (finite) operator $\cal O$;
suppose moreover that its renormalization was multiplicative and
its anomalous scaling therefore of the form
\eqn\dim{\mu{d\over d\mu}\O = -\gamma_\O (g^2)\O}
where $g^2$ is the coupling constant and $\mu$ is a renormalization
mass. (I will write all expressions
in such a way that they are naturally interpreted in dimensional
regularization with minimal or quasiminimal subtraction by taking $\mu$
to be the familiar scale factor of the extra dimensions, but
no firm commitment to that procedure is
necessary.) A finite multiplicative renormalization of $\cal O$
is a replacement of $\O$ by $\O'\equiv f(g^2) \O$, $f(g^2)$ being
a power series in $g^2$ with $f(0)=1$; the latter restriction
excludes redefinitions of the classical operator corresponding to
$\O$ that have nothing to do with the ambiguities of renormalized
perturbation theory. The anomalous dimension of $\O'$ differs
in general from that of $\O$ and is given by
\eqn\newdim{\gamma_{\O'}(g^2)=\gamma_{\O}(g^2) - \beta(g^2)
{f'(g^2)\over f(g^2)}.}
The power series for the beta-function starts at order $g^4$ and
that of $\gamma_{\O}$ typically at order $g^2$, but if $\gamma_{\O}$
has no one-loop term, $\gamma_{\O}=cg^4+...$, the quotient
$\gamma_{\O}/\beta$ is a power series in $g^2$
and a finite renormalization $f$ may be found such that the
renormalized operator $\O'$ has vanishing anomalous dimension,
\eqn\finren{f(g^2) = \exp \bigl[\int_0^{g^2} dx {\gamma_\O (x)\over
\beta(x)}\bigr].}
On the other hand, the part of the anomalous dimension that
arises at one loop is unaffected by finite renormalizations,
hence irremovable.
Similar reasoning applies to additive renormalizations; terms of
order $g^4$ and beyond in the power series expansion of an off-diagonal
anomalous dimension are removable by a finite renormalization, while the
one-loop, $g^2$ term is universal.

Anomalous Ward identities for the singlet axial current relate insertions
in Green functions of three operators: the divergence of $j^\mu_5$,
the singlet axial current itself; $j_5$, the pseudoscalar density; and the
anomalous term
$\ffd$. In addition they involve contact terms corresponding
to axial transformations of the fermion fields. To lowest nontrivial order
one finds the familiar expression
\eqn\ann{\dj=2mj_5 +{n_RT_Rg^2\over 16\pi^2}\FFd + c.t.,}
``$c.t.$'' being the contact terms. Upon insertion in a Green function
of fermion fields $\psi(x_1),...,\psi(x_p),$
$\bar\psi(y_1),...,\bar\psi(y_q)$
and gauge fields $A(z_1),...,A(z_r)$, ``$c.t.$'' takes the form
\eqn\ct{\eqalign{-\sum^p_{i=1} & \delta(w-x_i)
<\psi(x_1)...\gamma_5\psi(x_i)...
\psi(x_p)\bar\psi(y_1)...\bar\psi(y_q)A(z_1)...A(z_r)>\cr
& -\sum^q_{j=1}\delta(w-y_j)<\psi(x_1)...\psi(x_p)\bar\psi(y_1)...
\bar\psi(y_j)\gamma_5...\bar\psi(y_q)A(z_1)...A(z_r)>}}
involving the axial transformations of the fermions.

Classically the singlet axial current equals
$\bar\psi\gamma^\mu\gamma_5\psi$;
its quantum version is renormalized multiplicatively. The one-loop
diagram contributing to that (Figure 1) is identical to the diagram
for the vector current $\bar\psi\gamma^\mu\psi$ but for an
extra $\gamma_5$ at the vertex. The inclusion of this $\gamma_5$ does
not spoil the familiar cancellation of infinities between the
one-loop correction and the wave function renormalization of the
constituent fermion fields, so $j^\mu_5$, too, is finite at one loop
and has no one-loop anomalous dimension.
By the argument reviewed above one infers the existence of a
(unique) normalization that makes the scaling of the axial current
canonical to all orders; call this current $[j^\mu_5]_{C}$.
We may and will take the renormalized divergence of the axial current
to be the divergence of the renormalized axial current.

Similar reasoning applies to the pseudoscalar density $j_5$, classically
equal to $i\bar\psi\gamma_5\psi$. Its multiplicative renormalization
involves at one loop a diagram as in Figure 1. Now an analogy
obtains with the one-loop renormalization of $\bar\psi\psi$, the
renormalization constant of which is the inverse of that of the mass.
The combination $mj_5$ is thus found to scale canonically at one loop
and may be taken, via a finite renormalization of $j_5$ if necessary,
to have that property to all orders. I denote the resulting object
by $m[j_5]_{C}$.

The composite operator $\ffd\equiv \epsilon^{\mu\nu\rho\sigma}
F^a_{\mu\nu}F^a_{\rho\sigma}$ is
the derivative of the gauge-variant Chern-Simons current
$4\epsilon^{\mu\nu\rho\sigma}(A^a_\nu\partial_\rho A^a_\sigma + {1\over 3}
f_{abc}A^a_\nu A^b_\rho A^c_\sigma)$. The renormalization of this current
is not exhausted by a multiplicative factor, but involves mixing
with the axial fermion current as well. (By contrast, the axial current
itself is protected by its gauge invariance from non-diagonal
renormalization involving the gauge-variant Chern-Simons current.)
Correspondingly $\ffd$ mixes with $\dj$.

Figures 2 and 3 contain one-loop diagrams
relevant for the diagonal and off-diagonal renormalization of $\ffd$,
respectively. The analysis of Figure 2 is simplified
if background field Feynman rules are used, and I will do
so from now on. The logarithmically divergent parts of the diagrams in
Figure 2 then cancel among each other, so the diagonal, multiplicative
scaling behavior of $\ffd$ is derived entirely
(at least to this order) from the wave function renormalizations
of its constituent gluons. In the background field formalism
the renormalization of the gluon field is by construction exactly the
inverse of that of the coupling constant, so
the product $g^2\ffd$ is not renormalized and free of anomalous dimension
at one loop, again as far as diagonal renormalization is concerned.
The option of a finite
renormalization of $\ffd$ may then be invoked to extend this property to
all orders. A renormalized version of $\ffd$ will be denoted by $\ffdc$
if its product with $g^2$ has canonical diagonal scaling.

This leaves the issue of additive renormalization and
off-diagonal scaling, to be parametrized by
\eqn\ffdsc{\mu{d\over d\mu} \bigl(g^2\ffdc\bigr) =-\gamma_{F5}(g^2)\>g^2
\partial_\mu [j_5^\mu]_{C}.}
The extra $g^2$ on the right hand side balances the one on the left
in order that the counting of powers of $g^2$ in the anomalous
dimension $\gamma_{F5}$
be identical to that of the number of loops at which they arise.
Evaluation of the diagram in Figure 3 shows
$\gamma_{F5}$ to be nonvanishing at one loop; the logarithmically
divergent part of that diagram is of the form
\eqn\mix{-{3ic_Rg^2\over 4\pi^2}\ln{\Lambda^2\over\mu^2}
({\psl}'-\psl)\gamma_5}
in momentum space, $\Lambda$ being a cutoff and $\mu$ a renormalization
scale, and $p$ and $p'$ the momenta of the fermions. This implies
a one-loop value $\gamma_{F5}={3c_Rg^2/2\pi^2}$. Higher orders
may be changed by additive renormalizations of
$[\ffd]_C$, $([\ffd]_C)'\equiv [\ffd]_C +
f(g^2)\>\partial_\mu[j^\mu_5]_{C}$
with $f(0)=0$, but there is no particularly natural way of fixing this
freedom and I will not commit myself with regard to it, leaving the
ambiguity of $\ffdc$ implicit in the notation.

(There exists another approach to the normalization of $g^2\ffd$,
ultimately equivalent to the present one,
in which the Chern-Simons current is related by gauge or BRS
descent to a naturally finite operator, the normalization
of which fixes the diagonal normalization of the
renormalized $g^2\ffd$;
\nref\Bar{W.~Bardeen, Nucl. Phys. B75, 246 (1974).}
\nref\BMS{P.~Breitenlohner, D.~Maison and K.~Stelle, Phys.~Lett.
134B, 63 (1984).}
\nref\LPS{C.~Lucchesi, O.~Piguet and K.~Sibold, Int. J. Mod. Phys.
A2, 385 (1987).}see \refs{\Bar,\BMS,\LPS}.)

Now assume that the anomaly equation survives renormalization,
in the sense that there exists in the complete theory
a relation of linear dependence involving the renormalized operators
defined above and reducing to \ann\ at the lowest nontrivial order in
perturbation theory. Write this relation in the form
\eqn\anans{C_1(g^2)\>\partial_\mu[j^\mu_5]_{C} = C_2(g^2)\>2m[j_5]_C
+C_3(g^2)\>{n_RT_R\over 16\pi^2} g^2\ffdc + c.t..}
where the $C_i$ are power series in $g^2$, with $C_i(0)=1$ to
accommodate \ann, and the overall factor has been fixed by normalizing
the contact terms to their standard value, \ct.
Application of the renormalization group operator $\mu{d\over d\mu}$ to
both sides yields
\eqn\test{\beta\>C_1'\>\partial_\mu[j^\mu_5]_{C}=\beta\>C_2'\> 2m[j_5]_C+
\beta\>C_3'\>{n_RT_R\over 16\pi^2}g^2\ffdc
-C_3\>{n_RT_R\over 16\pi^2}g^2\gamma_{F5}\partial_\mu [j^\mu_5]_C.}
The contact terms have disappeared because they are
finite as they stand, see \ct, hence without anomalous dimension.
Without them the remaining operators are independent and consistency
requires
\eqn\testt{C_2'=C_3'=0,
\gap C_1'=-{n_RT_Rg^2\gamma_{F5}\over 16\pi^2\beta}C_3,}
i.e.
\eqn\testtt{C_1=1-{n_RT_R\over 16\pi^2}\int_0^{g^2} \!\!dx\>
{x\>\gamma_{F5}(x)\over\beta(x)}, \gap C_2=C_3=1.}
The single remaining non-universal feature (at least by the criteria
espoused in the above) may be hidden in notation by introducing
a finite renormalization of the canonical axial current,
\eqn\abj{[j^\mu_5]_{AB}\equiv C_1(g^2)[j_5^\mu]_C,}
in terms of which the renormalized anomaly equation reads
\eqn\anr{\partial_\mu [j^\mu_5]_{AB} = 2m[j_5]_C + {n_RT_R\over 16\pi^2}
g^2\ffdc + c.t.}
In keeping with the (mostly supersymmetric) literature $[j_5^\mu]_{AB}$
may be called an Adler-Bardeen current, with indefinite article to
signal the normalization ambiguity which it inherits,
via $C_1$, from the mixing dimension $\gamma_{F5}$. Along comes
a multiplicative anomalous dimension starting at order $g^4$
\nref\Cre{R.~Crewther, in {\it Facts and Prospects of Gauge Theories},
Acta Phys. Austr. Suppl. XIX, 47 (1978).}\refs{\AB,\Cre}.

\newsec{Dimensional Regularization}

The next step is to derive the anomaly in the framework
of dimensional regularization. This discussion is complementary to
the previous one in concentrating on aspects that are peculiar to a
specific
regularization scheme. For convenience I will mostly adhere to the
conventions and notation of Collins' textbook on renormalization
\ref\Col{J.~Collins, {\it Renormalization} (Cambridge University Press,
Cambridge U.K., 1984).}, where the Ward identity for the
(non-anomalous) non-singlet current is derived.

To define the axial current in the regulated theory
requires a definition for $\gamma_5$, which is notoriously problematic
in dimensional regularization. In the 't Hooft-Veltman
prescription \nref\BrM{P.~Breitenlohner and D.~Maison, Comm. Math.
Phys. 52, 11 (1977).}\refs{\tHV,\BrM}\ $\gamma_5$ is taken as the
product of the first four $\gamma$-matrices at the cost of
anticommutativity with all $\gamma$'s in $d$ dimensions. This requires
a distinction between the first four dimensions and the remaining $(d-4)$,
to which end one introduces projectors denoted by a superscripted bar
and hat, respectively; a vector $a^\mu$ is written as $\bar a^\mu
+ \hat a^\mu$, with $\bar a^\mu$ = $a^\mu$ if $\mu =0,1,2,3$ and
0 otherwise, and the other way around for $\hat a^\mu$. One may
then proceed to write down identities for tensorial objects with
contracted indices. The totally antisymmetric tensor
$\epsilon_{\mu\nu\rho\sigma}$ is introduced as a bar-carrying object
in the above sense, i.e. $\epsilon_{\mu\nu\rho\sigma}$
equals its usual value when each index is 0,1,2, or 3, and 0 otherwise;
for brevity of notation the bar will not be indicated explicitly.
Finally, $\gamma_5$ is defined by
\eqn\gamdef{\gamma_5 = {i\over 4!} \epsilon_{\mu\nu\rho\sigma}
\gamma^\mu\gamma^\nu\gamma^\rho\gamma^\sigma.}
Its most notable feature is its failure to anticommute with the hatted
$\gamma$'s:
\eqn\gamcom{\{\gamma^5, \bar\gamma^\mu\}=0 \>\>{\rm but}\>\>
[\gamma^5, \hat\gamma^\mu]=0.}
Trace identities may be established as usual: the trace of $\gamma_5$
times a product of $n$ $\gamma$-matrices vanishes if $n$ is odd or less
than four, while (taking $\tr\> {\bf 1}=4$)
\eqn\gamtr{\tr\> \gamma_5 \gamma^\mu \gamma^\nu \gamma^\rho
\gamma^\sigma= 4i\epsilon^{\mu\nu\rho\sigma}.}

The bare singlet axial current and pseudoscalar density in the
dimensionally regulated theory are now defined by
\eqn\barej{j^\mu_5 \equiv \bar\psi_0\bar\gamma^\mu\gamma_5\psi_0}
\eqn\bared{j_5 \equiv i\bar\psi_0\gamma_5\psi_0.}
$\psi_0 \equiv (Z_2)^{{1\over 2}}\psi$ is the bare fermion wave function.
The bar on $\gamma^\mu$ in the axial current, which does not affect
the classical limit, is necessary for hermiticity.

The would-be conservation law for the axial current is the
unrenormalized anomaly equation,
\eqn\anb{\partial_\mu j^\mu_5 = 2m_0j_5 + \Oan + \Oct}
where $\Oan$ and $\Oct$ are given by
\eqn\opan{\Oan\equiv {1\over 2} \bar\psi_0 \{\darr{\Dsl},\gamma_5\} \psi_0}
and
\eqn\opem{\Oct\equiv i\bar\psi_0\gamma_5 (i \Dsl -m_0)\psi_0
-i\bar\psi_0 (i\larr{\Dsl}+m_0)\gamma_5\psi_0.}
The operator $\Oan$ represents the potential anomaly. Only the hatted
components of the covariant derivation occur; $\Oan$ is ``evanescent''
\Col\
and would vanish if no divergences occurred in the theory in the limit
$d\rightarrow 4$. $\Oct$ is proportional to the equation of motion and
vanishes on shell, but its insertion in covariant Green functions,
where derivatives and time-ordering are taken to commute,
produces the contact terms associated with axial transformations
of the fermion as given explicitly by \ct.

The next step is to renormalize the equation, for which we adopt
minimal subtraction. The linearity of that procedure implies
that the subtracted equation is obtained by replacing each operator by
its minimally subtracted version (and $m_0$ by the minimally
renormalized mass $m$; we assume the coupling $g^2$ and wave function
renormalizations such as $Z_2$ to be defined by minimal subtraction
as well). The minimally subtracted current $[j^\mu_5]_{MS}$
and pseudoscalar density $[j_5]_{MS}$ are both multiplicatively related
to the respective unrenormalized operators. The contact terms are
manifestly finite, $[\Oct]_{MS}=\Oct$. These relations and the
anomaly equation implicitly determine how the minimally subtracted
anomaly operator $[\Oan]_{MS}$ is to be defined in terms of
unrenormalized operators and renormalization factors.

The crucial point is now that in the limit $d\to 4$ finite
contributions involving the subtracted anomaly term can only arise in
diagrams where the evanescent vertex on which that term is built
multiplies an internal divergence. As a result, $[\Oan]_{MS}$
survives in the limit $d\to 4$ as a linear combination of local
operators with non-evanescent basic vertices, the net effect being
a reduction in the number of independent operators.
The coefficients of this linear combination are (at least in perturbation
theory) power series in $g^2$ without constant terms, because $[\Oan]_{MS}$
vanishes at the tree level.
Other examples of this phenomenon have been worked out in contexts
rather different from the present one \nref\Bos{M.~Bos, Phys.~Lett.
189B, 435 (1987); Ann.~Phys.~(NY) 181, 177 (1988).}
\nref\Schu{C.~Schubert, Nucl. Phys. B323, 478 (1989).}\refs{\Bos,\Schu}.

The set of non-evanescent operators in terms of which $[\Oan]_{MS}$
is expressed may be restricted by dimensional and symmetry considerations.
That leaves $\partial_\mu[j^\mu_5]_{MS}$, $\mjms$, $\Oct$ and $\ffdms$,
the latter being the minimally subtracted version of the composite operator
$F\tilde F \equiv \epsilon^{\mu\nu\rho\sigma}(F^a_{\mu\nu})_0
(F^a_{\rho\sigma})_0$.
($(F^a_{\mu\nu})_0\equiv (Z_3)^{{1\over 2}}(F^a_{\mu\nu})_0$ is the
bare field strength, simply proportional to the renormalized one
in the background field method.) $\Oct$ may be replaced by other
operators by use of the minimally subtracted anomaly equation and the
remainder may be rearranged as
\eqn\opanr{[\Oan]_{MS} = f_1(g^2)\partial_\mu[j^\mu_5]_{MS}
+ f_2(g^2)m[j_5]_{MS} + f_3(g^2)\ffdms}
where the $f_i$ are all finite and of order $g^2$ at least.
Substitution in the
minimally subtracted anomaly equation yields
\eqn\and{(1-f_1(g^2))\partial_\mu[j^\mu_5]_{MS} =
(1+{\textstyle {1\over 2}}f_2(g^2))2m[j_5]_{MS} + f_3(g^2)\ffdms + \Oct.}
The existence of the anomaly thus amounts to $f_3\not= 0$.

The factors $1-f_1$ and $1+{\textstyle{1\over 2}}f_2$ may be removed by a
finite renormalization, leading to oversubtracted (with respect to
minimal subtraction, that is) operators:
\eqn\osj{[j^\mu_5]_{OS}\equiv (1-f_1(g^2))[j^\mu_5]_{MS}}
\eqn\osd{[j_5]_{OS}\equiv (1+{\textstyle{1\over 2}}f_2(g^2))[j_5]_{MS}.}
An analogous oversubtraction may be performed on $\ffd$; $f_3$, if
nonzero, has the form $f_3^{(1)}g^2 + \O(g^4)$ with
$f_3^{(1)}\not= 0$, so $f_3(g^2)/(f_3^{(1)} g^2)$ is a power
series starting with 1 and I renormalize
\eqn\osf{\ffdos\equiv {f_3(g^2)\over f_3^{(1)}g^2}\ffdms}
leading to a rewritten renormalized anomaly equation
\eqn\ancl{\djos = 2\mjos + f_3^{(1)}g^2\ffdos +\Oct.}

This brings the construction to the point where contact is made with
the discussion of the previous section. This time the renormalized
anomaly equation has been derived, rather than assumed.
Comparison with \anr\ identifies $[j_5^\mu]_{OS}$ as an
Adler-Bardeen current and $\mjos$ as the canonically scaling pseudoscalar
density $m[j_5]_C$; $g^2\ffdos$ is a product of the type
$g^2\ffdc$ with canonical diagonal scaling behavior.
The value of the off-diagonal anomalous dimension of $g^2\ffdos$
beyond one loop as well as the related anomalous dimension of
$[j_5^\mu]_{OS}$ emerge as specifics of the regularization.

\newsec{Calculations}

In this section I describe the calculation of quantities
relevant to the next-to-leading order two-gluon matrix element of
the anomaly equation.

The kinematical setting is as follows. We will compute
one-particle irreducible functions of two gauge fields with a single
operator insertion, that operator being one of those that occur in the
anomaly equation. The labeling of the external gauge fields will be as
in Figure 4, with $(a,b)$ and $(\mu,\nu)$ group and Lorentz indices,
respectively, and $p$ and $-p'$ the incoming gauge field momenta. For the
operators of interest $\O$ the two-gluon function
has the form
\eqn\mat{\epsilon_{\mu\nu\rho\sigma}p^\rho p'^\sigma \>\delta_{ab}
\>\Gh_{\O}(p,p';m)}
where $\Gh_{\O}$, to be referred to as the reduced matrix element,
is a function of the scalars $p^2,p'^2$ and $p\cdot p'$ as well
as of the fermion mass $m$. Where an expression for $\Gh_{\O}$ for
all $p$ and $p'$ is not practicable or illuminating I will specialize
to the infrared regime $p'\to p$, accessible from the complete
function after differentiation with respect to $p$, say.
I will then also eliminate either $p^2$ or $m^2$;
in each case the renormalized reduced matrix element, classically
dimensionless, will depend on the remaining parameter at most via
renormalization group logarithms. All computations are performed
using the background field Feynman rules, a convenient table
(and motivation) of which may be found in \ref\Abb{L.~Abbott,
Nucl. Phys. B185, 189 (1981).}.

The following ingredients are required for an evaluation of the
two-gluon matrix element of the anomaly equation
\ancl\ through next-to-leading order. The matrix
elements of $\djms$ and $\mjms$, which have their lowest nonzero
contributions at one loop, should be calculated at one and two loops.
The matrix element of $\ffdms$ starts at the tree level and we
need the one-loop contribution in addition to that. The functions
$f_1(g^2)$ and $f_2(g^2)$ contribute to the anomaly equation from the
next-to-leading level onward and their leading (i.e.$\>g^2$) terms
suffice. Of the function $f_3(g^2)$, on the other hand,
both the $g^2$ and the $g^4$ terms are required. The two-gluon matrix
element of $\Oct$ is zero.

We begin with the reduced matrix elements $\Gh_{\djms}$ and $\Gh_{\mjms}$.
The lowest order contributions stem from the familiar triangle graph,
Figure 4, and are easily computed for general $p$ and $p'$. The
results have been well-known since the early investigations of the anomaly,
\eqn\djone{\Gh_{\djms}(p,p';m)\1L=-{g^2n_RT_R\over \pi^2}\biggl({1\over 2}
+m^2I_{00}(p,p';m)\biggr)}
\eqn\mjone{\Gh_{\mjms} (p,p';m)\1L=-{g^2n_RT_R\over
2\pi^2}m^2I_{00}(p,p';m)}
where
\eqn\Ioo{I_{00}(p,p';m)\equiv \int^1_0 ds \int^{1-s}_0 \!\!dt
\>{1\over p^2t(1-t)+p'^2s(1-s)-2p\cdot p'st-m^2}.}
If the fermion mass vanishes,
\eqn\onemo{\Gh_\djms (p,p';m\is 0)\1L = -{g^2n_RT_R\over 2\pi^2},
\gap\Gh_{\mjms} (p,p';m\is 0) = 0;}
by contrast, if the fermion mass is finite but $p'\is p, p^2\to 0$
\eqn\onepo{\eqalign{\Gh_\djms& (p\is p',p^2\is 0;m)\1L = 0 \cr
\Gh_{\mjms} (p&\is p',p^2\is 0;m)\1L = {g^2n_RT_R\over 4\pi^2}.\cr}}

We proceed to the contributions of order $g^4$. The relation between
minimally subtracted and unrenormalized reduced matrix elements is
\eqn\ren{\Gh_{[\O]_{MS}} = Z_\O Z_A \Gh_{\O,bare}}
with $Z_\O$ the renormalization constant of $\O$ as determined by
minimal subtraction, $\O$ being any of the operators of interest;
$Z_A^{1\over 2}$ is the wave function renormalization of the gauge field.
For arbitrary $p,p'$ and $m$ the expressions for the reduced matrix
elements are rather unwieldy and I concentrate on the aforementioned
limits.

At the two-loop level $\Gh_{\dj,bare}$ and $\Gh_{j_5,bare}$
involve the eleven diagrams of Figure 5.
The values of these diagrams with
(1) $\O\is\dj$, $p\is p'$, $p^2\isnot 0$, $m\is0$, (2) $\O\is\dj$,
$p\is p'$, $p^2\to 0$, $m\isnot0$,
and (3) $\O\is j_5$, $p\is p'$, $p^2\to 0$, $m\isnot0$, are listed in
Tables 1, 2 and 3, respectively. They have been computed using the
background field Feynman rules. As we only determine terms through
order $g^4$ of the left hand side of \ren, the bare parameters
$g_0^2$ and $m_0$ have been replaced in the tables by the lowest order
of their renormalized expansions, $g^2\mu^{4-d}$ and $m$.
A common factor ${g^4 n_RT_R/(4\pi)^4}$ has been omitted; $L_m$ stands for
$\ln {m^2\over 4\pi\mu^2} + \gamma$ and $L_p$ for $\ln {-p^2\over
4\pi\mu^2} +\gamma$. The gauge dependence of the gluon
Feynman rules becomes an issue as one finds terms up to
quadratic order in the standard covariant gauge parameter $\alpha$.

Consider first $\dj$ for massless fermions. The two-loop diagrams of
Table 1 add up to
\eqn\djtwob{\eqalign{\Gh_{\dj,bare}& (p\is p',p^2\isnot 0;m\is 0)\2L
\sim\cr
&-{g^4n_RT_R\over 8\pi^4}
\Bigl(c_R+c_A\bigl(1+{3\over 4}(1-\alpha)-{1\over 8}(1-\alpha)^2
\bigr)\Bigr).\cr}}
The similarity sign is meant to express the replacement of bare parameters
by their lowest order renormalized ones.
The absence of a term proportional to $c_R\>\alpha$ corresponds to
the gauge invariance of the result for QED and obtains equally
from the other two tables. It is not hard to show by
diagrammatic reasoning that such terms must cancel independently of
the values of $p,p'$ or $m$. It may also be noted that the
propagator-induced corrections to the triangle, diagrams I through III,
vanish at $\alpha=0$, a consequence of the vanishing of the one-loop
self-energy of massless fermions in the Landau gauge. Proceeding to
the counterterms, the bare one-loop matrix element is
$-{g_0^2n_RT_R\over 2\pi^2} (p^2)^{(d-4)/2}$, as \onemo\ with the
regulating dimensionality $d$ taken into account;
but the bare coupling in this expression is converted to the finite
renormalized one by multiplication with $Z_A$ as in \ren, because
the renormalization of the wave function is precisely the inverse of
that of the coupling in the background field method. As for the
renormalization factor $Z_{\dj}$,
it is equal to 1 at one loop by the calculation sketched in section 2.
The renormalized reduced matrix element becomes, through two loops,
\eqn\djtwomsmo{\eqalign{\Gh_{\djms} &(p\is p',p^2\isnot 0;m\is 0)=
-{g^2n_RT_R\over 2\pi^2} \cr
& -{g^4n_RT_R\over 8\pi^4} \Bigl(c_R+c_A\bigl(1+{3\over 4}(1-\alpha)
-{1\over 8}(1-\alpha)^2\bigr)\Bigr).\cr}}

If the fermions are taken to be massive the two-loop unrenormalized
matrix element of $\dj$ is zero in the limit $p=p'$, $p^2\to 0$,
as is found by adding the entries of Table 2; as a result the
vanishing of the renormalized matrix element persists
to order $g^4$,
\eqn\djtwomspo{\Gh_{\djms}(p\is p',p^2\to 0;m\isnot 0)=0.}
This extends to the nonabelian case the corresponding property of
the divergence of the axial current in massive QED, proven to
all orders by Adler and Bardeen \AB; we shall return to this
presently.

Turning to the pseudoscalar density $j_5$, which occurs in the
anomaly equation multiplied by $m$ and is therefore relevant only in
the case of massive fermions, the contributions to the limit
$p\is p'$, $p^2\to 0$ tabulated in Table 3 add up to
\eqn\jtwob{\eqalign{&\Gh_{{j_5},bare}(p\is p',p^2\to 0;m\isnot 0)\2L
\sim\cr
&{g^4n_RT_R\over 16\pi^4m}
\Bigl({1\over 2}c_R+c_A\bigl(1+{3\over 4}(1-\alpha)-{1\over 8}(1-\alpha)^2
\bigr)\Bigr).\cr}}
Note that though the infrared behavior of individual diagrams
produces terms involving logarithms of $p^2$, such terms cancel in
the sum.
The bare one-loop result is here
\eqn\joneb{\Gh_{{j_5},bare}(p\is p',p^2\to 0;m\isnot 0)\1L =
{g_0^2n_RT_R\over 4\pi^2m_0}(m_0)^{d-4}.}
The counterterms implicit in $g_0^2$ cancel against those generated
by $Z_A$, as before. A similar cancellation arises between $1\over m_0$
and $Z_{{j_5}}$; we saw in section 2 that the one-loop renormalization
of $j_5$ is opposite to that of $m$, $m_0j_5=m[j_5]_{MS}$. That
leaves only the scale factor, which by use of the familiar minimal
subtraction for the mass
\eqn\mren{m_0=(1-{3g^2c_R\over 8\pi^2\epsilon}+\ldots)\>m}
is found to contribute $3g^4c_Rn_RT_R/ 32\pi^4m$ to the reduced
matrix element. We obtain, to two loops,
\eqn\jtwoms{\eqalign{\Gh_{{[j_5]_{MS}}} &(p\is p',p^2\to 0;m\isnot0)=
{g^2n_RT_R\over 4\pi^2m}\cr
&+{g^4n_RT_R\over 16\pi^4m}
\Bigl(2c_R+c_A\bigl(1+{3\over 4}(1-\alpha) -{1\over 8}(1-\alpha)^2
\bigr)\Bigr).\cr}}
Thus $\Gh_{2\mjms}$ for massive fermions is similar, but not completely
identical, to $\Gh_{\djms}$ for massless ones.

The results above may also be arrived at by subtracting subdivergences
diagram by diagram. This allows for the special case of the Feynman gauge
comparison with some of the ordinary perturbation theory
results of \Akh, which are presented in this form; I have found
agreement wherever I checked. Upon further restriction to massless
fermions one also finds agreement with the older results of \JL.
Restriction both to massless fermions and to the group
structure $c_R$ produces a value given in \ref\Gab{G.~Gabadadze and
A.~Pivovarov, JETP Letters 54, 298 (1991).}.

The next task is to find the functions $f_1(g^2)$ and $f_2(g^2)$ at order
$g^2$. This may be done by computing the one-loop matrix element of the
anomalous operator $\Oan$ with a fermion and a conjugate fermion field;
the tree graph vanishes in the limit $d\to 4$. The relevant one-loop
diagrams are shown in Figure 6. After inclusion of the wave
function renormalization of the constituent fermions of $\Oan$
one finds the finite result, in a general gauge $\alpha$,
\eqn\opanone{ {ig^2c_R\over 4\pi^2} ({\bar\psl}'-{\bar\psl})\gamma_5
-{ig^2c_R\over \pi^2} m\gamma_5. }
This is rewritten in the form of the right hand side of \opanr\ by
comparison with the basic vertices of $\dj$ and $j_5$, equal to
$i({\bar\psl}'-\bar\psl)\gamma_5$ and $i\gamma_5$, respectively, and
observing that the $\psi$-$\bar\psi$ matrix element of $\ffd$
starts only at order $g^2$. It follows that
\eqn\fonetwo{f_1(g^2)={g^2c_R\over 4\pi^2} +\ldots,\gap
f_2(g^2)=-{g^2c_R\over \pi^2} +\ldots}
Obviously this is basically an abelian effect, hence independent of
$\alpha$.

The above results may now be combined into two-loop expressions
for the reduced matrix elements of the oversubtracted fermionic operators
$\djos\equiv (1-f_1)\djms$ and $\jos\equiv
(1+{\textstyle{1\over 2}}f_2)\jms$ in a general covariant gauge $\alpha$:
\eqn\djtwoosmo{\eqalign{\Gh_{\djos} &(p\is p',p^2\isnot 0;m\is 0)=
-{g^2n_RT_R\over 2\pi^2} \cr
&-{g^4c_An_RT_R\over 8\pi^4}\Bigl(1+{3\over 4}(1-\alpha)
-{1\over 8}(1-\alpha)^2\Bigr)\cr}}
\eqn\djtwoospo{\Gh_{\djos} (p\is p', p^2\to 0;m\isnot 0)=0.}
\eqn\jtwoospo{\eqalign{\Gh_{m\jos} &(p\is p',p^2\to 0;m\isnot 0)=
{g^2n_RT_R\over 4\pi^2} \cr
&+{g^4c_An_RT_R\over 16\pi^4}
\Bigl(1+{3\over 4}(1-\alpha) -{1\over 8}(1-\alpha)^2\Bigr)\cr}}
The finite renormalization has had the effect of eliminating terms
involving the quadratic Casimir of the fermion representation
from the $g^4$ contributions.

We now turn to the matrix element of $\ffdms$. The tree level value,
determined by the basic vertex of $\ffd$, is
$\Gh_\ffdms\vert_{tree} =-8$.
The next order is obtained from a formula of the form \ren.
The unsubtracted one-loop diagrams are shown in Figure 2; in
accordance with the considerations of section 2, their sum is finite
if $g_0^2$ is replaced by $g^2\mu^{4-d}$. The renormalization
constant $Z_\ffd$ must therefore be chosen to compensate for the
wave function renormalizations of the gluon fields in $\ffd$,
$Z_\ffd=(Z_A)^{-1}$.
With this choice, $\Gh_\ffdms$ at one loop is found to be,
for arbitrary $p,p'$,
\eqn\ffdone{\eqalign{\Gh_\ffdms\1L &= -{2g^2c_A\over \pi^2}
\biggl\{1-{1\over 2}(p-p')^2I_{00}(p,p';0)\cr
+{1\over 4}(1-\alpha)\Bigl[1+\int_0^1\!ds\int_0^{1-s}&\!\!\!dt
\>{1-s-t\over\Delta}\bigl((1-2s)p^2+(1-2t)p'^2+2(s+t)p\cdot p'\bigr)\cr
+(p-p')^2\int_0^1\!ds &\int_0^{1-s}\!\!\!dt \>{1-s-t\over\Delta^2}
\bigl(t(1-s)p^2+s(1-t)p'^2)\bigr)\Bigr]\cr
&\gap -{1\over 8}(1-\alpha)^2\biggr\}\cr}}
with $I_{00}$ as in \Ioo\ and
\eqn\denom{\Delta \equiv p^2t(1-t)+p'^2s(1-s)-2p\cdot p'st.}
This somewhat unsavory expression has an easily read off limit as
$p'\to p$; including tree and one-loop effects,
\eqn\ffdlim{\Gh_\ffdms(p'=p) = -8\Bigl(1+{g^2c_A\over 4\pi^2}
\bigl(1+{3\over 4}(1-\alpha)
-{1\over 8}(1-\alpha)^2\bigr)\Bigr).}
Note that to this order the fermions play no role in $\Gh_\ffdms$.

The one-loop matrix element of $\ffd$ was computed using a different
intermediate regularization in \ref\Ans{A.A.~Anselm and A.A.~Jogansen,
Sov. Phys. JETP 69, 670 (1989).}. There the term proportional to
$(1-\alpha)$ is presented as
\eqn\ffdans{-{c_Ag^2\over \pi^2}(1-\alpha)\bigl[1-{1\over 4}(p-p')^2
\bigl(p^2{\partial\over \partial p^2} +p'^2
{\partial\over\partial p'^2}\bigr)I_{00}\bigr]}
where the partial derivatives are defined by treating
$I_{00}(p,p';0)$ as a function of the invariants $p^2$, $p'^2$ and
$p\cdot p'$. This appears to yield ${1\over 2}(1-\alpha)$ instead
of ${3\over 4}(1-\alpha)$ in \ffdlim\ \Jog, but
$I_{00}$ is singular as $p'\to p$,
\eqn\Ioolim{\eqalign{&I_{00}(p,p';0) \to -{4\over (p+p')^2}\ln
{(p'-p)^2\over (p'+p)^2}\cr
&\bigl(p^2{\partial\over \partial p^2} +p'^2
{\partial\over\partial p'^2}\bigr)\>I_{00} \to
-{2\over (p'-p)^2}\cr}}
and this eliminates the apparent discrepancy (cf. also
\SVII).

With an eye on \ancl\ we turn now to the determination of $f_3^{(1)}$, the
Taylor coefficient of $g^2$ in $f_3(g^2)$. As with $f_1$ and $f_2$,
the $g^2$ term in $f_3$ may be extracted from a one-loop diagram,
in this case the triangle with two gluons (Figure 4). With $\Oan$ at
the upper vertex this graph yields, in the limit $d\to 4$ (to
eliminate evanescent contributions) \tHV\
\eqn\opanaaone{\Gh_{\Oan} (p,p';m)\1L = -{g^2n_RT_R\over 2\pi^2}.}
This is compatible with \opanr\ provided
\eqn\fthree{f_3(g^2)={g^2n_RT_R\over 16\pi^2}+\ldots,}
as the first two terms on the right hand side of \opanr\
contribute to $\Gh_{\Oan}$ only at the $g^4$ level.
$f_3^{(1)}$ is thus determined to be $n_RT_R/16\pi^2$, and
one recognizes in it the well-known coefficient of the anomaly.

We may now verify this by evaluating the two-gluon matrix element
of the renormalized anomaly equation \ancl\ using the results obtained
above. There are no contact terms. The reduced matrix element of
$\djos-2m\jos$ was found to be the same in the two kinematical
settings studied through two loops,
\eqn\djtwotot{\eqalign{\Gh_{\djos} -2\> \Gh_{m\jos}&=
-{g^2n_RT_R\over 2\pi^2} \cr
&-{g^4c_An_RT_R\over 8\pi^4}\bigl(1+{3\over 4}(1-\alpha)
-{1\over 8}(1-\alpha)^2\bigr).\cr}}
But this is exactly equal to $f_3^{(1)}\>\Gh_\ffdms$ in the
limit $p'\to p$, so the validity of the
renormalized anomaly equation requires that
there be no oversubtraction for $\ffd$, i.e., that
$\ffdms$ and $\ffdos$ be the same object.

The verification of this consistency requirement, which previous
treatments \refs{\JL,\Akh} have omitted altogether, occupies the
remainder of this section. One way of accomplishing it would be to
directly compute the term of order $g^4$ in $f_3$ that gives
the finite renormalization relating $\ffdms$ to $\ffdos$ at second
order. I will forego this option for a less direct and therefore
maybe more interesting check.
$\ffdos$ is characterized among the finite multiplicative
renormalizations of $\ffdms$ by the property that $g^2\ffdos$
has no anomalous diagonal scaling. Ergo, $\ffdms$ is equal to
$\ffdos$ if and only if its diagonal anomalous dimension is
exactly $\beta(g^2)/g^2$.

For this to be the case the minimal subtraction constant $Z_\ffd$
must be equal to $Z_A^{-1}$ through two loops; we already know that to
be true at one loop. In view of \ren\ this amounts to
$\Gh_{\ffd,bare}$ being finite if its parameters are expanded
to order $g^4$.

The thirty-six two-loop diagrams contributing to $\Gh_{\ffd,bare}$
are shown in Figure 7. Diagrams in which both external gluons couple
to the same four-point vertex may be discarded immediately by
symmetry considerations and have been omitted; no orientations
are indicated on fermion and ghost lines and it is understood that
both choices are to be considered. Table 4
lists the pole parts of these diagrams in the Feynman gauge
($\alpha = 1$) as computed for $p'\to p$; for the diagrams
involving fermions, the limits $p^2/m^2\to 0$ and $m^2/p^2\to 0$
are distinguished if necessary. Each contribution
of the form $1\over\epsilon^2$ is to be supplemented by a
$-{1\over\epsilon}L_p$ term of equal strength, left out of the table
for the sake of brevity.

The contributions of order $g^4$ carry three distinct
group theoretical structures, namely $c_A^2$, $c_An_RT_R$, and
$c_Rn_RT_R$. The cancellation of infinities we are after should
obtain for each structure separately.

Consider first the $c_A^2$ terms. Retaining only those terms is
tantamount to assuming that the fermions are absent from the theory,
which reduces the set of diagrams to numbers I through XXIX.
The sum of these diagrams as per Table 4 is
\eqn\ffdtwob{\Gh_{\ffd,bare}\2L \sim -{59c_A^2g^4\over 48\pi^4\epsilon}.}
The cancellation of $1\over \epsilon^2$
poles in this sum, with the accompanying cancellation of terms of the
form ${1\over\epsilon}L_p$, reflects the absence of divergences in the
one-loop
matrix element \ref\tHII{G.~'t~Hooft, Nucl. Phys. B61, 455 (1973).}.
The counterterms follow from expanding
$g_0^2$ and the bare gauge parameter $\alpha_0$ in $\Gh_{\ffd,bare}\1L$
in renormalized quantities; if the beta-function is written as
$b_1g^4+\ldots$ and the renormalization group coefficient
of the gauge parameter, $\mu{d\over d\mu}\alpha$, as
$d_1g^2+\ldots$ , that amounts to applying
\eqn\ffdct{{b_1g^4\over\epsilon} {\partial\over\partial g^2} +
{d_1g^2\over\epsilon} {\partial\over\partial\alpha}}
to the expression in \ffdlim. With the well-known values
$b_1=-{11c_A/24\pi^2}$ and $d_1={5c_A\alpha/24\pi^2}$
that leads in the Feynman gauge to a counterterm contribution
exactly opposite to \ffdtwob. This shows that the $g^4$ part of
$\Gh_{\ffd,bare}$ is finite and hence to this order in pure QCD
$Z_\ffd=Z_A^{-1}$, as desired.

Upon re-inclusion of the fermions there are seven more two-loop diagrams,
each involving a single fermion loop, shown as numbers XXX through
XXXVI in Figure 7. The Feynman gauge pole parts of these diagrams,
again computed at $p'\is p$, are sensitive to the value of $p^2/m^2$.
Adding up the relevant entries of Table 4 produces
\eqn\ffdfer{\delta\Gh_{\ffd,bare}\2L
\sim{7g^4 c_An_RT_R\over 12\pi^4\epsilon} +
\cases{{\displaystyle{{3g^4c_Rn_RT_R\over 4\pi^4\epsilon}}}&if
$m^2/p^2\to 0$;\cr
0&if $p^2/m^2\to 0$.\cr}}
Concomitantly the values of $b_1$ and $d_1$ receive the
well-known fermionic contributions $\delta b_1 = n_RT_R/6\pi^2$
and $\delta d_1 = -n_RT_R\alpha/6\pi^2$.
It is easy to check that the pole terms with group structure $c_AT_R$
cancel between the new diagrams and the ($\alpha\is 1$)
counterterm modifications. Thus it only remains to account for
the $c_Rn_RT_R$ part of \ffdfer.

Inspection of Table 4 shows that the net $c_Rn_RT_R$ contribution arises
from diagram XXXI. The upper part of that diagram is identical to the
one-loop mixing diagram of $\ffd$ and $\dj$, Figure 3. In fact it is
not $\ffd$ but $\ffd+{3c_Rg^2\over 2\pi^2\epsilon}\dj$ that is
multiplicatively related to $\ffdms$, where the pole is that of
the mixing diagram; this formula accounts for \mix,
with $\ln{\Lambda\over\mu}$ replaced by $1\over\epsilon$. The
renormalization formula for $\ffd$, \ren, should be modified accordingly.
In view of the one-loop values of $\Gh_{\dj}(p'=p)$ in the two limiting
regimes of $p^2/m^2$, (4.5-6), this effect provides a correction
that precisely cancels the $c_Rn_RT_R$-dependent term in \ffdfer.
The peculiar dependence of \ffdfer\ on the
presence or absence of a fermion mass is thus seen to be inherited
from the analogous feature of the divergence of the axial current.

This concludes the consistency check (at least as far as the
Feynman gauge is concerned): through two loops
the minimally subtracted operator $\ffdms$ scales as the inverse
of the minimally subtracted coupling constant, without the need
for a finite renormalization.

\newsec{Comments}

In the previous section the two-gluon matrix elements of a number of
suitably normalized operators were computed through two loops in order
to verify the validity of the renormalized anomaly equation.
I will now consider those matrix elements in their own right.

The first issue that should be addressed is the nature of the various
kinematical limits. Most of the explicit results listed above
have applied to the reduced matrix elements of composite
operators at total momentum $p'-p$ equal to zero, leaving the momentum
squared of each external gauge field, $p^2$, and the mass $m$ of the
fermions as independent parameters. Existing physical applications
typically involve the reduced matrix elements in the limits in which
$-p^2$ goes either to zero or to infinity with respect to $m^2$.
It is well-known that at one loop the reduced matrix element of the
current divergence vanishes as $-p^2/m^2\to 0$, whereas the pseudoscalar
density $mj_5$ drops out in the opposite limit, see (4.5-6). Evidently
neither of these observations depends on the precise normalization of the
objects they apply to. Both statements remain true at two loops, the first,
exhibited in \djtwomspo, being obviously the non-trivial one.
Such behavior has
been observed in the formalism of \Akh\ as well. The noncommutativity of the
infrared limits has also been discussed at the two-loop level in a
dispersive approach in \ref\Zak{V.~Zakharov, Phys. Rev. D42, 1208 (1990).};
however, the contention of that paper that different limits correspond
to different currents is at odds with the point of view taken here.

For the abelian theory the vanishing to all orders of the reduced
matrix element of $\dj$ in the limit $-p^2/m^2\to 0$ is an aspect
of the original Adler-Bardeen analysis \AB; the proof uses gauge invariance
and analyticity. The background field method transfers
at least the former tool to the nonabelian case. A putative generalized
proof would start with the observation that if symmetry considerations
are taken into account, the 1PI function of the (renormalized) current
$j_5^\lambda$ and two gauge fields has the form \ref\Ros{L.~Rosenberg,
Phys. Rev. 129, 2786 (1963).}, at ${p'}^2\is p^2$,
\eqn\john{\eqalign{\Bigl[ D_1(p^2)\bigl( \epsilon^\lambda{}_{\!\mu\nu\alpha}
\>p^\alpha
&+\epsilon^\lambda{}_{\!\mu\nu\alpha}\>{p'}^\alpha\bigr)
+ D_2(p^2)\bigl( p_\mu\epsilon^\lambda{}_{\!\nu\alpha\beta}p^\alpha
{p'}^\beta + p'_\nu\epsilon^\lambda{}_{\!\mu\alpha\beta}
p^\alpha {p'}^\beta\bigr) \cr
+& D_3(p^2)\bigl( p_\nu\epsilon^\lambda{}_{\!\mu\alpha\beta}p^\alpha
{p'}^\beta
+p'_\mu\epsilon^\lambda{}_{\!\nu\alpha\beta}p^\alpha {p'}^\beta\bigr)\Bigr]
\delta_{ab}\cr}}
and the (renormalized) reduced matrix element of $\partial_\lambda
j^\lambda_5$ is $-2iD_1$.
Gauge invariance of the background field effective action
requires \john\ to vanish when contracted with
$p^\mu$ or ${p'}^\nu$, implying $D_1(p^2)=-p^2\bigl(
D_2(p^2) + D_3(p^2) \bigr)$. If $D_2$ and $D_3$ are sufficiently
well-behaved at $p^2=0$ it follows that $D_1(p^2=0)=0$; but
the premise is not at all obvious in the nonabelian case, whereas
the $D$'s are known to be analytic near $p^2=0$ in the abelian theory with
massive fermions.

The interpolation between the two extremes $-p^2/m^2\to 0$
and $m^2/(-p^2)\to 0$ involves $m$ explicitly and is therefore
highly scheme-dependent. At low orders, however, relatively simple
results may emerge; for instance, for the two-loop $(1-\alpha)^2$ terms of
$\Gh_{\djms}$ and $\Gh_{\mjms}$ the interpolation turns out to
exactly parallel that of the complete one-loop functions, as in
\djone\ and \mjone.

Granted the distribution of nonzero results between the
divergence of the current and the pseudoscalar density,
what are those results taken by themselves?
The simplest case is the massive abelian one at $p\is p'$, $p^2\to 0$,
treated by Adler and Bardeen \AB.
In that case the anomaly equation relates the matrix element of the
canonical pseudoscalar density $m[j_5]_C$ to that of $g^2\ffdc$;
the latter is no longer sensitive to mixing because the matrix
element of $\dj$ vanishes under the circumstances. (In this sense
what is often referred to in the literature as the Adler-Bardeen
current is merely an accessory to the Adler-Bardeen \AB\ theorem.)
It can be shown by inspection of diagrams that the complete reduced
matrix element of $g^2\ffdc$, hence also that of $m[j_5]_C$, is
given by its lowest order value, a constant times $g^2$. Each of
these two reduced matrix
elements satisfies a renormalization group equation
\eqn\peter{\Bigl[\mu{\partial\over\partial\mu} +
\beta{\partial\over\partial g^2}
+ \gamma_m m^2{\partial\over\partial m^2} + \beta_\alpha
{\partial\over\partial\alpha}
- 2\gamma_A\Bigr] \Gh_{\O} (p'\is p,p^2\is 0;m\isnot 0) =0}
where the anomalous dimension of the photon is
essentially identical to the beta-function, $\gamma_A \is
\beta/2g^2$; the exact results for the $\Gh_{\O} (p'\is p;
p^2\is 0;m\isnot 0)$
thus satisfy the renormalization group equations \peter\
in the minimal nontrivial way \Zee,
with $\mu{\partial\over\partial\mu}\Gh_{\O}=0$.
The explicit two-loop values, \jtwoospo\ and \ffdlim\ at $c_A\is 0$,
are obviously compatible with the general result.

Still in the abelian theory but with vanishing fermion mass,
it becomes impossible to disentangle the divergence of the axial
current and the antisymmetric
photon tensor by the infrared behavior of their reduced
matrix elements, because of mixing and the attendant normalization
ambiguity. Perhaps the cleanest matrix element in this kinematical
configuration is that of the divergence of the canonical axial
current, $\partial_\mu [j^\mu_5]_C$, which satisfies a renormalization
equation of the form \peter. However, the vanishing anomalous dimension
of $[j^\mu_5]_C$ now merely conceals the underlying non-trivial
renormalization effects and the solution to \peter\ is no longer exhausted
by a one-loop term. Explicitly, from \abj, \testtt, \mix, and
\djtwoosmo\ (with $c_A=0$), and using the lowest order value of $\beta$,
\eqn\fred{\Gh_{{\partial_\mu [j^\mu_5]_C}} (p'\is p; m\is 0)\vert_{abelian}
= -{g^2n_RT_R\over 2\pi^2}
-{9g^4c_Rn_RT_R\over 32\pi^4} +\ldots.}
The renormalization group calls for a logarithmic term at order $g^6$
to match the $g^4$ contribution; this term is supplied by three-loop
diagrams as in Figure 8, computed and discussed in \Ans.
Of course higher powers of $\ln {-p^2\over\mu^2}$ will emerge
from the ellipsis if the calculation is pursued to higher orders.

Turning now to the nonabelian case, consider only the case of massive
fermions and $p'\is p, p^2\to 0$, obviously the simpler. In
ordinary perturbation theory the analogy with the abelian situation is
obscured from the onset, because the anomalous dimension of the gluon,
$\gamma_A$, is then no longer simply related to the beta-function;
this spoils the chances of a purely lowest order reduced matrix element,
as this could no longer saturate the renormalization
group equation. In the background field method, $\gamma_A$ is
again equal to ${1\over 2}{\beta\over g^2}$.
As a result there can be no logarithms in the reduced matrix
element of $m[j_5]_C$ or $g^2\ffdc$ at the next-to-leading order,
a fact borne out by the explicit results, \jtwoospo\ and \ffdlim\
(in contrast to the corresponding quantities in \Akh, obtained
in ordinary perturbation theory). Yet, as these equations show, even in the
background field formalism $\Gh_{m[j_5]_{C}} (p'\is p, p^2\to 0;m\isnot 0)$
and $\Gh_{g^2\ffdc} (p'\is p,p^2\to 0;m\isnot 0)$ have corrections to their
leading orders, in contrast to the abelian case.

A salient feature of these corrections is their dependence on the
gauge parameter $\alpha$. That an expression for general $p$ and $p'$
such as \ffdone\ exhibits such dependence is not surprising, because
there is no reason to expect gauge invariant
results as long as the external gluons are off-shell. Of course
infrared problems make it impossible to take $p^2\to 0$ or ${p'}^2\to 0$
in \ffdone. If $p'$ is first taken to be equal to $p$ the one-loop result
becomes independent of $p^2$ and $p'^2$, see \ffdlim, and in particular
finite as $p^2\to 0$, but this is deceptive; the vanishing of the anomalous
dimension of $g^2\ffdc$ is accommodated in $\Gh_{g^2\ffdc} (p'=p)$
by the occurrence of powers of $\ln{-p^2\over\mu^2}$ starting at the
two-loop level.

The noncommutativity of these limits is
reminiscent of what happens with the divergence of the axial
current at one loop. The analogy may be reinforced by introduction
of a small finite gluon mass $M$. This was probably first
considered by Chanowitz \ref\Cha{M.~Chanowitz,
Phys. Rev. D9, 503 (1974).}, who employed spontaneous symmetry breaking
to procure a mass for the gauge fields. As long as gauge fields contribute
through no more than one loop the mass may also be introduced
straightforwardly, via a massive propagator
\eqn\mike{-i\Bigl[{\eta_{\mu\nu}\over k^2-M^2}-{(1-\alpha)k^\mu k^\nu\over
(k^2-M^2)(k^2-\alpha M^2)}\Bigr].}
Recalculation of the diagrams in Figure 2 with this propagator
leads to a generalization of \ffdone. In the Feynman gauge,
\eqn\charles{\Gh_\ffdms\1L = -{2g^2c_A\over \pi^2}\Bigl(1+
\bigl(2M^2-{1\over 2}(p-p')^2\bigr)I_{00}(p,p';M)\Bigr).}
This should be compared with the expression for the one-loop reduced
matrix element of $\dj$, \djone. At $p'\is p$, i.e. as the composite
operator is integrated over spacetime, the two results are
completely analogous. In particular, as long as $M$ is finite
the one-loop correction to $\Gh_\ffdms$ disappears at $p'\is p$. This
is actually also true for gauges $\alpha\isnot 1$, though the expression
for general $p$ and $p'$ is not quite as simple as \charles\ above.
The impossibility of a direct generalization of the Adler-Bardeen
results to the nonabelian case appears therefore attendant upon
the masslessness of the gauge fields.

We may reconsider similarly the calculation of $\Gh_{{[j_5]_{MS}}}
(p\is p',p^2\to 0;m\isnot 0)$ by including a small gluon mass, taking
the limit ${M\over m}\rightarrow 0$.
Not surprisingly, the entries in Table 3 that betray their sensitivity
to the infrared behavior of the gauge fields by the presence of
$\ln (-p^2)$ have to be changed to reflect the new circumstances. Diagrams
I through VI, VIII and XI are not affected, but VII now yields
\eqn\ben{c_A\Bigl((-12-4\alpha)\ln{M^2\over m^2}
-4\alpha\ln\alpha+2+6\alpha\Bigr)}
and IX+X
\eqn\jerry{2c_A\Bigl( {4\alpha\over\epsilon} -4\alpha L_m+
(6+2\alpha)\ln{M^2\over m^2} +2\alpha\ln\alpha -2-4\alpha\Bigr).}
Compared with table 3 as it stands the extra contributions add up
to $-16c_A\bigl[1+{3\over 4}(1-\alpha)-{1\over 8}(1-\alpha)^2\bigr]$,
which is precisely what is needed to eliminate the terms proportional
to $c_Ag^4$.

The existence of a finite one-loop contribution to the reduced matrix
element of $g^2\ffdc$ at $p'=p$ appears to be at odds with a
semiclassical interpretation of ${1\over 64\pi^2}g^2\ffdc$ as the
Pontryagin density, the integral of which
is the instanton number. It is tempting to try to remove the extra factor
by a finite renormalization, but that would interfere with the vanishing
of the anomalous dimension of $g^2\ffdc$ and leave explicit logarithms
untouched.
However, the sensitivity of the one-loop reduced matrix element to the
infrared arrangement of the theory suggests that the problem is
an infrared artifact of perturbation theory, and that its vanishing at
infinitesimal gluon mass mimicks behavior induced by a nonperturbative
physical infrared cutoff. (This appears to be also the point of view
advocated in \Jog; the alternative point of view that all or part of
the one-loop matrix element of $g^2\ffd$ is a signature of a real effect
involving the vacuum angle $\theta_{QCD}$ \ref\Vac{R.~Jackiw and
C.~Rebbi, Phys. Rev. Lett. 37, 172 (1976); C.~Callan, R.~Dashen and
D.~Gross, Phys. Lett. 63B, 334 (1976).}
is developed in \refs{\SVI,\SVII}.)

\newsec{Conclusion}

Since its identification more than twenty years ago, the axial anomaly
has never been far from the spotlight of theoretical physics. Regarding
most of its aspects it could therefore be plausibly argued that
no relevant insight remains to be put forward. Yet the continuing flow of
papers related to the interplay of the anomaly and the process of
renormalization bears witness that the concrete
organization of the currently available intuition in specific
physical contexts is still not without challenges. The rationale of
the present paper was the consideration that for nonabelian gauge
theory and the practically important formalism of dimensional
regularization, a truly comprehensive discussion was lacking in
the literature.

A review was given of the renormalization group arguments that
lead to a convenient normalization of the singlet axial current
and the operators related to its divergence.
These considerations were matched
to a derivation of the anomaly equation using dimensional
regularization with what is probably the most straightforward
treatment of pseudotensorial objects, that according to 't Hooft and
Veltman. This settled the issue in principle. To illustrate how this
works in practice, and verify that the abstract argument was sound,
I computed the two-gluon matrix element of each term in
the anomaly equation through two loops, including features
neglected in previous treatments. Such calculations supply,
apart from the desired reassurance, matrix elements that
may be of independent interest, if not of immediate physical
significance.

\vfill
\eject
\centerline{{\bf Acknowledgements}}\nobreak
\bigskip
This work was supported in part by the National Science Foundation
under grant NSF/PHY-89-15286.
I have benefited from discussions with members of the UCLA Theoretical
Physics Group, particularly Brian Hill, and with Stefano Forte.

% The references
\listrefs

% The tables
\centerline{{\bf Table Captions}}
\bigskip
\item{Table 1. }
Two-loop contributions to the unrenormalized reduced matrix element of
$\dj$ at $p'\is p$ and $m\is 0$, numbered as in Fig. 5.
A common factor $g^4n_RT_R/(4\pi)^4$ is omitted.
\smallskip

\item{Table 2. }
Two-loop contributions to the unrenormalized reduced matrix element of
$\dj$ for $p'\is p$, $p^2\to 0$ and $m\isnot 0$, numbered as in Fig. 5.
A common factor $g^4n_RT_R/(4\pi)^4$ is omitted.
\smallskip

\item{Table 3. }
Two-loop contributions to the unrenormalized reduced matrix element of
$j_5$ for $p'\is p$, $p^2\to 0$ and $m\isnot 0$, numbered as in Fig. 5.
A common factor $g^4n_RT_R/(4\pi)^4m$ is omitted.
\smallskip

\item{Table 4. }
Pole parts of the two-loop Feynman gauge contributions to the
unrenormalized reduced matrix element of $\ffd$ for $p'=p$, numbered
as in Fig. 7. A common factor $g^4/(4\pi)^4$ is
omitted, as are all terms of the form ${1\over\epsilon}L_p$.
In XXXI, (*) marks the result for $m\is 0$, $p^2\isnot 0$, (**) that
of $m\isnot 0$, $p^2\to 0$.

\bigskip
\bigskip
\bigskip

% The figures
\centerline{{\bf Figure Captions}}
\bigskip
\item{Fig. 1. }
One-loop diagram relevant for the renormalization of $j_5^\mu$
and $j_5$.
\smallskip
\item{Fig. 2. }
One loop diagrams contributing to the two-gluon 1PI function of
$\ffd$.
\smallskip
\item{Fig. 3. }
One-loop diagram leading to mixing of $\dj$ with $\ffd$.
\smallskip
\item{Fig. 4. }
One-loop diagram for the two-gluon 1PI functions of $\dj$,
$j_5$, or $\Oan$.
\smallskip
\item{Fig. 5. }
Two-loop diagrams contributing to the two-gluon 1PI functions
of $\dj$ and $j_5$.
\smallskip
\item{Fig. 6. }
One-loop diagrams contributing to the $\psi$-$\bar\psi$
1PI function of $\Oan$.
\smallskip
\item{Fig. 7. }
Two-loop diagrams contributing to the two-gluon 1PI function
of $\ffd$. For each ghost or quark line both orientations should be
taken into account.
\smallskip
\item{Fig. 8. }
A three-loop contribution to the two-gluon 1PI function of $\dj$.
\vfill
\eject

\hoffset .2truein
\hsize 6.5truein
\def\cR{{\rm c}_{\rm R}}
\def\cA{{\rm c}_{\rm A}}
\def\TR{{\rm n}_{\rm R}{\rm T}_{\rm R}}
\def\Lm{{{\rm L}_m}}
\def\Lp{{{\rm L}_p}}
\def\tablerule{\noalign{\medskip}}
\def\interline{\noalign{\vskip 5pt plus 2pt minus 2pt}}
\def\ee{{\epsilon}}
\def\EE{{\epsilon^2}}
\nopagenumbers

\centerline{{\bf TABLE 1}}
\bigskip
$$\vbox{\halign{\bf#\hfil&\quad\hfil {$\displaystyle{#}$}\hfil&\quad
\quad\hfil {$\displaystyle{#}$}\hfil& \quad\hfil {$\displaystyle{#}$}\hfil\cr
$\,$ & (1-\alpha)^0 & (1-\alpha)^1 & (1-\alpha)^2 \cr
\noalign{\medskip\hrule\medskip}
\tablerule
I + II & 2\cR \Bigl( {12\over\epsilon} -12\Lp +21\Bigr)
& 2\cR \Bigl( -{12\over\epsilon} +12\Lp -21\Bigr)
& 0\cr
\tablerule
III & \cR \Bigl( {8\over\epsilon} -8\Lp +14\Bigr)
& \cR \Bigl( -{8\over\epsilon} +8\Lp -14\Bigr)
& 0\cr
\tablerule
IV+V & 2(\cR -{\textstyle{1\over 2}}\cA)\Bigl( -{8\over\epsilon}
+8\Lp -18\Bigr)
& 2(\cR-{\textstyle{1\over 2}} \cA) \Bigl( {8\over\epsilon} -8\Lp +14\Bigr)
& 0\cr
\tablerule
VI & \cR \Bigl( -{16\over\epsilon} +16\Lp -52\Bigr)
& \cR \Bigl( {16\over\epsilon} -16\Lp +28\Bigr)
& 0\cr
\tablerule
VII & \cA \Bigl( -{32\over\epsilon} +32\Lp -104\Bigr)
& \cA \Bigl( {8\over\epsilon}-8\Lp +10\Bigr)
& 0\cr
\tablerule
VIII & 0 & 0 & 0\cr
\tablerule
IX+X & 2\cA \Bigl( {4\over\epsilon} -4\Lp +{31\over 3} +8\zeta (3)\Bigr)
& 2\cA \Bigl( {2\over\epsilon} -2\Lp -{9\over 2} -4\zeta (3)\Bigr)
& 4\cA\cr
\tablerule
XI & \cA \Bigl( {16\over\epsilon} -16\Lp + {100\over 3} - 16\zeta(3)\Bigr)
& \cA \Bigl( -{4\over\epsilon} + 4\Lp -11 +8\zeta(3)\Bigr)
& 0\cr
\tablerule
\noalign{\medskip\hrule\medskip}
total & -32\cR-32\cA & -24\cA & 4\cA\cr}}$$
\vfill
\eject

\centerline{{\bf TABLE 2}}
\bigskip
$$\vbox{\halign{\bf#\hfil&\quad\hfil {$\displaystyle{#}$}
\hfil&\quad\quad\hfil {$\displaystyle{#}$}\hfil&
\quad\hfil {$\displaystyle{#}$}\hfil\cr
$\,$ & (1-\alpha)^0 & (1-\alpha)^1 & (1-\alpha)^2\cr
\noalign{\medskip\hrule\medskip}
\tablerule
I + II & 2\cR \Bigl( {4\over\epsilon} -4\Lm -{11\over 3}\Bigr)
& 2\cR \Bigl( {4\over\epsilon} -4\Lm -3\Bigr)
& 0\cr
\tablerule
III & \cR \Bigl( -{24\over\epsilon} +24\Lm + {10\over 3}\Bigr)
& \cR \Bigl( {8\over\epsilon} -8\Lm -2\Bigr)
& 0\cr
\tablerule
IV+V & 2(\cR -{\textstyle{1\over 2}} \cA)\Bigl(
{8\over\epsilon} -8\Lm +2\Bigr)
& 2(\cR-{\textstyle{1\over 2}} \cA) \Bigl( -{8\over\epsilon} +8\Lm +2\Bigr)
& 0\cr
\tablerule
VI & 0
& 4\cR
& 0\cr
\tablerule
VII & \cA \Bigl( -{32\over\epsilon} +32\Lm +8\Bigr)
& \cA \Bigl( {8\over\epsilon}-8\Lm -2\Bigr)
& 0\cr
\tablerule
VIII & 0 & 0 & 0\cr
\tablerule
IX+X & 2\cA \Bigl( {12\over\epsilon} -12\Lm -1\Bigr)
& 2\cA \Bigl( -{6\over\epsilon} +6\Lm +{5\over 2} \Bigr)
& 0\cr
\tablerule
XI & \cA \Bigl( {16\over\epsilon} -16\Lm -4\Bigr)
& \cA \Bigl( -{4\over\epsilon} + 4\Lm -1\Bigr)
& 0\cr
\tablerule
\noalign{\medskip\hrule\medskip}
total & 0 & 0 & 0 \cr}}$$
\vfill
\eject

\centerline{{\bf TABLE 3}}
\bigskip
$$\vbox{\halign{\bf#\hfil&\quad\hfil {$\displaystyle{#}$}\hfil&\quad
\quad\hfil {$\displaystyle{#}$}\hfil& \quad\hfil {$\displaystyle{#}$}\hfil\cr
$\,$ & (1-\alpha)^0 & (1-\alpha)^1 & (1-\alpha)^2 \cr
\noalign{\medskip\hrule\medskip}
\tablerule
I + II & 2\cR \Bigl( {-16\over\epsilon} +16\Lm -{16\over 3}\Bigr)
& 2\cR \Bigl( {8\over\epsilon} -8\Lm -2 \Bigr)
& 0\cr
\tablerule
III & \cR \Bigl( -{16\over\epsilon} +16\Lm-{16\over 3}\Bigr)
& \cR \Bigl( {8\over\epsilon}-8\Lm-2\Bigr)
& 0\cr
\tablerule
IV+V & 2(\cR -{\textstyle{1\over 2}} \cA)\Bigl( {8\over\epsilon}-8\Lm+2\Bigr)
& 2(\cR-{\textstyle{1\over 2}} \cA) \Bigl( -{8\over\epsilon}+8\Lm+2\Bigr)
& 0\cr
\tablerule
VI & \cR \Bigl( {32\over\epsilon}-32\Lm+20\Bigr)
& \cR \Bigl( -{8\over\epsilon}+8\Lm+2\Bigr)
& 0\cr
\tablerule
VII & \cA \Bigl( -16\Lp+16\Lm+56\Bigr)
& \cA \Bigl( 4\Lp-4\Lm-6\Bigr)
& 0\cr
\tablerule
VIII & 0 & 0 & 0\cr
\tablerule
IX+X & 2\cA \Bigl( {4\over\epsilon} +8\Lp -12\Lm -22\Bigr)
& 2\cA \Bigl( -{4\over\epsilon} -2\Lp +6\Lm +10\Bigr)
& -2\cA\cr
\tablerule
XI & 6 \cA & 0 & 0\cr
\tablerule
\noalign{\medskip\hrule\medskip}
total & 8\cR+16\cA & 12\cA & -2\cA \cr}}$$
\vfill
\eject

\centerline{{\bf TABLE 4}}
$$\vbox{\halign{\bf#\hfil&\quad\hfil {$\displaystyle{#}$}
\hfil&\quad \quad \quad {\bf#}\hfil&
\quad\hfil {$\displaystyle{#}$}\hfil\cr
\noalign{\hrule\medskip}
I & \Bigl( -{148\over\EE} -{541\over \ee}\Bigr)\cA^2
& XIX & 0\cr
\interline
II & \Bigl( -{320\over\EE} -{1104\over\ee}\Bigr) \cA^2
& XX & {48\over\ee}\cA^2\cr
\interline
III & \Bigl( {16\over\EE} +{2\over\ee}\Bigr) \cA^2
& XXI & \Bigl( {192\over\EE} +{624\over\ee}\Bigr)\cA^2\cr
\interline
IV & \Bigl( {12\over\EE} +{35\over \ee}\Bigr)\cA^2
& XXII & \Bigl( -{192\over\EE} -{480\over\ee}\Bigr) \cA^2\cr
\interline
V & 0 & XXIII & \Bigl( {108\over \EE}+{255\over \ee}\Bigr)\cA^2\cr
\interline
VI & \Bigl( -{2\over\ee}\Bigr)\cA^2
& XXIV & \Bigl( -{4\over\EE} -{9\over \ee}\Bigr)\cA^2\cr
\interline
VII & \Bigl( -{608\over 3\EE} - {5960\over 9\ee}\Bigr)\cA^2
& XXV & \Bigl( {144\over\EE} +{956\over 3\ee}\Bigr)\cA^2\cr
\interline
VIII & \Bigl( -{304\over 3\EE} -{2716\over 9\ee}\Bigr)\cA^2
& XXVI & \Bigl( {16\over\EE} +{140\over 3\ee}\Bigr)\cA^2\cr
\interline
IX & \Bigl( -{32\over 3\EE} -{344\over 9\ee}\Bigr)\cA^2
& XXVII & \Bigl( -{192\over\EE} -{384\over \ee}\Bigr)\cA^2\cr
\interline
X & \Bigl( -{16\over 3\EE} -{196\over 9\ee}\Bigr)\cA^2
& XXVIII & \Bigl( -{36\over\EE} -{81\over \ee}\Bigr)\cA^2\cr
\interline
XI & -{18\over\ee}\cA^2
& XXIX & \Bigl( -{4\over\EE} -{9\over \ee}\Bigr)\cA^2\cr
\interline
XII & {2\over\ee}\cA^2
& XXX & \Bigl( -{192\over \EE} -{432\over\ee}\Bigr)\cA\TR\cr
\interline
XIII & \Bigl( {144\over\EE} + {396\over\ee}\Bigr)\cA^2
& XXXI & {128\over\ee}\cR\TR\>(*), -{64\over\ee}\cR\TR\>(**)\cr
\interline
XIV & 0
& XXXII & {64\over\ee}(\cR-{\textstyle{1\over 2}}\cA)\TR\cr
\interline
XV & \Bigl( {108\over\EE} + {243\over \ee}\Bigr)\cA^2
& XXXIII & \Bigl( {512\over 3\EE} + {4736\over 9\ee}\Bigr)\cA\TR\cr
\interline
XVI & \Bigl( -{4\over\EE} -{1\over \ee}\Bigr)\cA^2
& XXXIV & \Bigl( {256\over 3\EE} +{2176\over 9\ee}\Bigr)\cA\TR\cr
\interline
XVII & \Bigl( {192\over\EE} + {576\over\ee}\Bigr)\cA^2
& XXXV & \Bigl( {64\over\EE} + {112\over\ee})\cA\TR\cr
\interline
XVIII & \Bigl( {288\over\EE} + {792\over\ee}\Bigr)\cA^2
& XXXVI & \Bigl( -{128\over \EE} - {800\over 3\ee}\Bigr)\cA\TR\cr
\noalign{\medskip\hrule}}}$$
\vfill
\eject
\end